\documentclass[12pt]{article}
\usepackage{amsmath,amssymb,epsfig}
\begin{document}
\parindent=0pt
\parskip=6pt
\rm

\vspace*{0.5cm}

\begin{center}
{\bf \Large Novel results about magnetic fluctuation effects near the
normal-to-superconducting phase transition in a zero magnetic field}

\vspace{0.3cm}
 D. V. SHOPOVA$^{\ast}$, T. P. TODOROV$^{\dag}$

{\em  CPCM Laboratory, Institute of Solid State Physics,\\
 Bulgarian Academy of Sciences, BG-1784 Sofia, Bulgaria.} \\
\end{center}

$^{\ast}$ Corresponding author: sho@issp.bas.bg

$^{\dag}$ Permanent address: Joint Technical College at the Technical
University of Sofia.

\vspace{1cm}


\vspace{0.2cm}


\vspace{0.2cm}

\begin{abstract}

A systematic treatment of the magnetic fluctuations effect on the
properties of the normal-to-superconducting phase transition in a zero
external magnetic field is given within the self-consistent
approximation and the quasi-macroscopic Ginzburg-Landau model. New
results for thin superconducting films are presented. Thermodynamic
quantities having a direct experimental interest  as the order
parameter jump, latent heat, and specific heat are considered and
numerically evaluated for bulk Al and thin Al films. The possibility
for an experimental verification of the theoretical predictions is
discussed.
\end{abstract}

\section{Introduction}
\label{sec.I.}

In 1974 Halperin, Lubensky and Ma (HLM)~\cite{Halperin:1974} showed
that the magnetic fluctuations change the order of the superconducting
phase transition in a zero external magnetic field ($H_0 = |\vec{H}_0|
= 0)$), i.e., the order of the phase transition from normal-to-uniform
(Meissner) superconducting state at $T_{c0} = T_c(H_0 = 0)$.  In the
mean-field approximation, when both magnetic and superconducting
fluctuations are neglected, this phase transition is of second order;
see, e.g., Refs.~\cite{Lifshitz:1980,Uzunov:1993}. Moreover, the
fluctuations $\psi(\vec{x}) = [\psi(\vec{x}) - \langle
\psi(\vec{x})\rangle]$ of the superconducting order parameter
$\psi(\vec{x})$ towards the statistical average $\langle
\psi(\vec{x})\rangle$ are extremely small and can be safely ignored in
usual low-temperature ($T_{c0} < 20$ K) superconductors. For a long
time these superconductors have been considered as an excellent example
of a standard phase transition of second order described by the
mean-field approximation. When the magnetic fluctuations are taken into
account in the Ginzburg-Landau (GL) free energy $F(\psi, \vec{A})$ of
superconductor~\cite{Lifshitz:1980}, the same normal-to-superconducting
phase transition in a zero (mean) external magnetic field $(\vec{H}_0 =
0)$ is found to be  a weakly-first order phase transition with a very
small latent heat which cannot be observed by available experimental
techniques~\cite{Halperin:1974}. The effect of a magnetic fluctuation
change of the  superconducting phase transition order, called HLM
effect, is very weak in bulk (three dimensional, or 3D) superconductors
even in
 Al where the GL number $\kappa$ is very small ($\kappa \ll
1$) - a circumstance which is in favor of the
effect~\cite{Halperin:1974,Chen:1978}.

In this paper we shall investigate this fluctuation-induced first order
phase transition in thin (quasi-2D) superconducting films. Bulk
superconductors will  be also discussed in order to compare them with
the behavior of thin films. We shall use a self-consistent
approximation~\cite{Halperin:1974}, in which the fluctuations
$\delta\psi$ of $\psi$ are neglected but the magnetic fluctuations are
completely taken into account. Note, that the so-called tree
approximation~\cite{Uzunov:1993} does not yield the HLM effect and the
self-consistent, or mean-field-like, approximation, mentioned above, is
the simplest analytical method for an investigation of this phenomenon.

The present study is intended to provide enough theoretical results
about the behavior of measurable physical quantities directly related
to the phase transition properties and in this way to ensure a
theoretical basis for future experiments on the existence of the HLM
effect. The need of an experimental observation of the HLM effect is
very important because the effect remains a theoretical paradigm
without a reliable experimental verification although its mechanism
 - the interaction of gauge fields in a quite universal
Abelian-Higgs model - is of fundamental interest for different fields
of physics as pure~\cite{Lawrie:1982, Folk:2001, Shopova:2002,
Shopova1:2003, Shopova2:2003, Shopova3:2003}, and
disordered~\cite{Uzunov:1983, Athorne:1985, Busiello:1986,
Blagoeva:1990, Busiello:1991, Uzunov:1990, Fisher:1988, Fisher1:1990}
superconductors, quantum phase transitions~\cite{Shopova4:2003,
Kirkpatrick:2003}, scalar electrodynamics~\cite{Coleman:1973}, liquid
crystals~\cite{Gennes:1993, Lubensky:1978, Anisimov:1990, Garland:1990,
Yethiraj:2000}, and cosmology~\cite{Linde:1979, Vilenkin:1994}. On the
other hand, there are some theoretical studies, based on Monte Carlo
simulations~\cite{Bartholomew:1983}, the so-called dual
model~\cite{Kiometzis:1994, Kiometzis:1995}, and certain variants of
the renormalization-group (RG)~\cite{Nogueira:2000,Kleinert:2001}, in
which no evidence of HLM effect was reported; for a discussion of this
point, see, the review article~\cite{Folk:1999}. Therefore, in the
modern theory of phase transitions the problem for the existence of HLM
effect is controversial and cannot be easily solved without a hint from
the experiment. The experimental research of the effect in liquid
crystals cannot be considered reliable although the reported results
are in favor of its existence.

Recently, it has been shown~\cite{Folk:2001} that the HLM effect is
stronger in quasi-2D superconducting films than in bulk superconductors
and the preliminary evaluation of the relevant physical quantities like
the order parameter jump and the latent heat at the equilibrium point
of the fluctuation-induced first order transition in superconducting
films gives for them several orders bigger values than for those in
bulk
materials~\cite{Shopova:2002,Shopova1:2003,Shopova2:2003,Shopova3:2003}.
This result reopens the problem for an experimental search of HLM
effect in type I superconductors, in particular, in thin films of type
I superconductors with relatively small GL parameter $\kappa$. Here we
shall investigate this problem in a comprehensive way.

We shall neglect the fluctuations of the superconducting order
parameter because their effect on the thermodynamics of the
superconductor is very weak; see, e.g.,
Refs.~\cite{Lifshitz:1980,Uzunov:1993}. Within this approximation, the
problems in the scope of the work can be considered without the use of
RG, as well as of numerous and quite interesting RG results available
in the literature; for a review, see, e.g., Ref.~\cite{Folk:1999}.

In Sec.~2 we present a  derivation of the effective free energy of a
D-dimensional superconductor. In Sec.~3 we give the first thorough
investigation of the effective free energy for  bulk superconductors.
In Sec.~4  the quasi-2D superconducting films and the validity of the
Landau expansion are discussed.  In Sec.~5 we summarize our main
conclusions.

\section{EFFECTIVE FREE ENERGY}
\label{sec.II.}

\subsection{Model considerations}
\label{sec.2.1.}

The GL free energy~\cite{Lifshitz:1980} of a D-dimensional
superconductor of volume $V_D = (L_1...L_D)$ is given in the form
\begin{equation}
\label{eq1}
 F(\psi,\vec{A}) = \int d^D x \left[ a|\psi|^2 + \frac{b}{2}|\psi|^4 +
\frac{\hbar^2}{4m}\left|\left(\nabla - \frac{2ie}{\hbar
c}\vec{A}\right)\psi \right|^2 + \frac{\vec{B}^2}{8\pi}\right]\:.
\end{equation}
In Eq.~(\ref{eq1}) the first Landau parameter $a = \alpha_0(T-T_{c0})$
is expressed by the critical temperature $T_{c0}=T_c(H=0)$ in a zero
external magnetic field ($H=|\vec{H}|$), $b
> 0$ is the second Landau parameter and $e \equiv |e|$ is the electron charge. The square $B^2$ of the
magnetic induction $\vec{B} = (\vec{H} + 4\pi\vec{M})$, is given by the
vector potential $\vec{A}(\vec{x})=\{ A_j(\vec{x}), \; j=1,...,D\}$ in
the form
\begin{equation}
\label{eq2}
 \vec{B}^2 = \frac{1}{2} \sum^{D}_{i,\:j\:=\:1} \left( \frac{\partial A_j}{\partial x_i} -
 \frac{\partial A_i}{\partial x_j}\right)^2\:,
\end{equation}
here the vector potential $\vec{A}(\vec{x})$ obeys the Coulomb gauge
$\nabla \cdot \vec{A}(\vec{x}) = 0$. For a 3D superconductor the
relation $\vec{B} = [\nabla \cdot \vec{A}(\vec{x})]$ can be used  and
when $\vec{B}=\vec{B}_0$ is uniform along the $z$- axis, the Landau
gauge $\vec{A}_0(\vec{x}) = B_0 (-y/2,-x/2,0)$ can be applied. This
representation can be generalized for $D >2$ - dimensional systems,
where the magnetic induction $B_0$ is a second rank tensor:
\begin{equation}
\label{eq3}
B_{0ij} =B_0 (\delta_{i1}\delta_{j2} -
\delta_{j2}\delta_{i1}).
\end{equation}

If we use the notation $\vec{x} = (x_1,x_2,\vec{r})$, where $\vec{r}$
is a $(D-2)$ - dimensional vector perpendicular to the plane
$(x_1,x_2)$, in the $3D$ case we shall have $\vec{r}=(0,0,z)$, and

\begin{equation}
\label{eq4}
B_j = \frac{1}{2}\epsilon_{jkl}B_{0kl} = B_0\delta_{j3}\;,
\end{equation}
where $\epsilon_{jkl}$ is the antisymmetric Levi-Civita symbol. The
Landau gauge and Eqs.~(\ref{eq3})~-~(\ref{eq4}) can be used for uniform
$\vec{B}=\vec{B}_0$ when $\delta \vec{B}$ - fluctuations are neglected;
see, e.g. Ref.~\cite{Brezin:1985}. In the prevailing part of our study
we shall apply the general  Coulomb gauge of the field
$\vec{A}(\vec{x})$ which does not exclude spatial dependent magnetic
fluctuations $\delta \vec{B}(\vec{x})$.

In nonmagnetic superconductors where the mean value
$\langle\vec{M}\rangle =(\vec{M}-\delta\vec{M})$ of the magnetization
$\vec{M}$ is equal to zero in the normal state in zero external
magnetic field, the magnetic induction in presence of external magnetic
field takes the form:
\begin{equation}
\label{eq5} \vec{B} = \vec{H}_0 + \delta \vec{H}(\vec{x}) +4\pi\delta
\vec{M}(\vec{x})\;,
\end{equation}
where $\vec{H}_0$ is the (uniform) regular part of the external
magnetic field and $ \delta \vec{H}$ is an irregular part of $\vec{H}$
created by uncontrollable  effects. We neglect the irregular part
$\delta \vec{H}$ and set $\vec{H}_0=0$, then $\vec{B}$ contains only a
fluctuation part $ \vec{B} \equiv \delta \vec{B}(\vec{x}) = 4\pi\delta
\vec{M}(\vec{x})$ that describes the diamagnetic variations of
$\vec{M}(\vec{x})$ around the zero value $\langle\vec{M}\rangle =0$ due
to fluctuations $\delta\psi(\vec{x})$ of the ordering field
$\psi(\vec{x})$ above $(T>T_{c0})$ and below $(T<T_{c0})$ the
normal-to-superconducting transition at $T_{c0}$. Note, that the
non-fluctuation  part $\vec{A}_0=[\vec{A}(\vec{x})-
\delta\vec{A}(\vec{x})]$ corresponds to the regular part $\vec{B}_0 =
(\vec{H}_0 + \langle \vec{M}\rangle) = 0$ of $\vec{B}$ in nonmagnetic
superconductors $(\langle \vec{M}\rangle = 0)$ in a zero external
magnetic field $(\vec{H}_0 =0)$. Then we can set $\vec{A}_0(\vec{x})=0$
and, hence, $\delta\vec{A}(\vec{x})=\vec{A}(\vec{x})$, so we have an
entirely fluctuation vector potential $\vec{A}(\vec{x})$ which
interacts with the order parameter $\psi(\vec{x})$. This interaction
can be of type $\psi^2A$ and  $\psi^2A^2$ and generates all effects
discussed in the paper.

We accept periodic boundary conditions for the superconductor surface.
This means to ignore the surface energy including the additional energy
due to the penetration of the magnetic field in a surface layer of
thickness equal to the London penetration depth
$\lambda(T)=\lambda_0|t_0|^{-1/2}, \; t_0= |T-T_{c0}|/T_{c0}$;
$\lambda_0= (mc^2b/8\pi e^2\alpha_0 T_{c0})^{1/2}$ is the
zero-temperature value of $\lambda$. This approximation is adequate for
superconductors of thickness $L_0 \gg \lambda(T)\gg a_0$, where  $a_0$
is the lattice constant and $L_0 = min\{L_i,\: i=1,...,D \}$. As we
suppose the external magnetic field to be zero $(H_0=0) $ or very small
in real experiments, the requirement $L_0 \gg \lambda(T)$ cannot be
satisfied and we take into account only the condition $L_0 \gg a_0$.

In microscopic models of periodic structures the periodic boundary
conditions confine the wave vectors $\vec{k}_i= \{k_i=(2\pi n_i/L_i);\;
i=1,...,D\}$ in the first Brillouin zone $[ -(\pi/a_0)\le k_i <
(\pi/a_0)]$ and the expansion of their values beyond this zone can be
made either by neglecting the periodicity of the crystal structure or
on the basis of the assumption  that big wave numbers $k=|\vec{k}|$
have a negligible contribution to the calculated quantities. The last
argument is widely accepted in the  phase transitions theory where the
long-wavelength $(ka_0\ll 1)$ limit can be used. In particular, this
argument is valid in the continuum limit $(V_D/a_0^D \to \infty)$.
Therefore, for both crystal and nonperiodic structures we can use a
cutoff $\Lambda \sim (\pi/a_0)$ and afterwards to extend this cutoff to
infinity provided the main contributions in the summations over
$\vec{k}$ come from the relatively small wavenumbers $(k \ll \Lambda)$.
Note, that here we make a quasimacroscopic description based on the GL
functional~(\ref{eq1}) which means that the microscopic phenomena are
excluded from our consideration.

The GL free energy functional takes into account phenomena with
characteristic lengths $\xi_0$ and $\lambda_0$ or larger ($\xi$ and
$\lambda$) where $\lambda(T)$ is the London penetration length
mentioned above and $\xi(T) = \xi_0|t|^{-1/2}$ is the coherence
length~\cite{Lifshitz:1980}; here $\xi_0 =(\hbar^2/4m \alpha_0
T_{c0})^{1/2}$ is the zero-temperature coherence length. In
low-temperature superconductors $\xi_0$ and $\lambda_0$ are much bigger
than the lattice constant $a_0$.  Having in mind this argument we shall
assume that in our investigation $\Lambda \ll (\pi/a_0)$. Whether the
upper cutoff $\Lambda$ is chosen to be either $\Lambda \sim 1/\xi_0$ or
$\Lambda \sim 1/\lambda_0$ is a problem that has to be solved by
additional arguments (see Sec.~3.3).

We shall use the Fourier expansion
\begin{equation}
\label{eq6} A_j(\vec{x}) = \frac{1}{V_D^{\frac{1}{2}}} \sum_k
A_j(\vec{k})e^{i \vec{k}. \vec{x}}
\end{equation}
and
\begin{equation}
\label{eq7} \psi (\vec{x}) = \frac{1}{V_D^{\frac{1}{2}}} \sum_k \psi
(\vec{k})e^{i \vec{k}. \vec{x}}\; ,
\end{equation}
where the Fourier amplitudes $A_j(\vec{k})$ obey the relation
$A_j^{\ast}(\vec{k}) = A_j(- \vec{k})$ and  $\vec{k} . \vec{A}(\vec{k})
= 0$. The Fourier amplitude  $\psi (\vec{k})$ is not equal to
$\psi^{\ast} (- \vec{k})$ because $\psi (\vec{x})$ is a complex
function. For the same reason $\psi(0) \equiv \psi (\vec{k} =0)$ is a
complex number.

\subsection{Approximations}
\label{2.2.}

The total ignoring of both superconducting and magnetic fluctuations in
Eq.~(\ref{eq1}) leads to the familiar free approximation where the GL
equations~\cite{Lifshitz:1980} should be solved. Note, that the free,
or mean-field, approximation is the lowest order theory within the
framework of the loop expansion, e.g., see
~\cite{Uzunov:1993,Zinn-Justin:1993}. The systematic treatment of the
fluctuation effects in the asymptotic vicinity of the phase transition
point can be given by RG.

The effect of the superconducting fluctuations $\delta \psi (\vec{x})$
on the phase transition properties is restricted in a negligibly small
vicinity $(|T-T_{c0}| \sim 10^{-12} \div 10^{-16})$ of the temperature
$T_{c0}$ and we shall assume that $\delta \psi (\vec{x}) = 0$, i.e.,
$\psi \approx \langle \psi (\vec{x}) \rangle$; from now on we shall
denote $\langle \psi (\vec{x}) \rangle$ by $\psi$. So we apply a
mean-field approximation with respect to the order parameter $ \psi
(\vec{x})$. Within this approximation we shall take into account the
$\delta \vec{A}(\vec{x})$-fluctuations  for $\vec{B}_0 = 0$, i.e.,
$\vec{A}(\vec{x})= \delta \vec{A}(\vec{x})$.  Furthermore, the
$\vec{A}(\vec{x})$-fluctuations can be integrated out from the
partition function, defined by:
\begin{equation}
\label{eq8} Z(\psi) = \int {\cal{D}}A e^{-F(\psi, \vec{A})/k_B T} \;,
\end{equation}
where the functional integral $ \int {\cal{D}}A$ is defined by
\begin{equation}
\label{eq9} \int^{\infty}_ {-\infty} \prod_{j=1}^D \prod_{x \in V_D} d
A_j(\vec{x}) \delta[\mbox{div} \vec{A} (\vec{x})] \;.
\end{equation}
The integration is over all possible configurations of the field
$\vec{A}(\vec{x})$; the $\delta$-function takes into account the
Coulomb gauge.

The partition function $Z(\psi)$ corresponds to an effective free
energy $\cal{F}$
\begin{equation}
\label{eq10} {\cal{F}}_D =  - k_B T \ln{Z(\psi)} \;,
\end{equation}
The magnetic fluctuations will be completely taken into account, if
only we are able to solve exactly the integral~(\ref{eq8}). The exact
solution can be done for a uniform order parameter $\psi$. The uniform
value of $\psi$ is different from the mean-field value of $\psi$
because the uniform fluctuations of  $\psi(\vec{x})$ always exist, so
we should choose one of these two possibilities. The problem for this
choice arises after the calculation the integral~(\ref{eq8}) at a next
stage of consideration when the effective free energy $\cal{F}_D$ is
analyzed and the properties of the superconducting phase $(\psi > 0)$
are investigated. The effective free energy is a particular case of the
effective thermodynamic potential in the phase transition
theory~\cite{Uzunov:1993,Zinn-Justin:1993} and we must treat the
uniform $\psi$ in the way prescribed in the field theory of phase
transitions. It will become obvious from the next discussion that we
shall use a loop-like expansion which can be exactly summed up to give
a logarithmic dependence on $|\psi|^2$.

Because of the spontaneous symmetry breaking of the continuous symmetry
in the ground state, the ordered phase $\psi
> 0$, i.e., the effective free energies discussed in this paper depend on the
modulus $|\psi|$ of the complex number $\psi = |\psi| e^{i\theta}$ but
not on the phase angle $\theta$ which remains arbitrary. That is why we
shall consider the modulus $|\psi|$ as an ``effective order parameter"
because the angle $\theta$ does not play any role in the phenomena
investigated in the paper. The quantity $|\psi|$ remains undetermined
up to the stage when we define the equilibrium order parameter
$|\psi_0|$ by the equation of state $[\partial
{\cal{F}}_D(\psi)/\partial \psi] = 0$. This equation gives the
equilibrium value $\psi_0$ of $\psi$ and the difference  $\delta \psi_0
= (\psi_0$ - $\psi)$ can be treated as the uniform (zero dimensional)
fluctuation of the field $\psi(\vec{x})$. The $\vec{x}$-dependent
fluctuations $\delta \psi (\vec{x})$ have been neglected because of the
uniformity of $\psi$. The solution $\psi_0$ will be stable towards the
uniform fluctuation $\delta \psi$ provided the same solution $\psi_0 =
|\psi_0|e^{i\theta_0}$ corresponds to a stable (normal or
superconducting) phase; the phase angle $\theta_0$ remains unspecified.
 Therefore, we begin our
investigation setting $\psi$ uniform but at some stage of consideration
we shall also ignore the uniform fluctuation $\delta \psi$ and deal
only with the equilibrium value $\psi_0$ of $\psi$. The equilibrium
value will be calculated after taking into account magnetic
fluctuations, so it will be different from the usual result $|\psi_0| =
(|a|/b)^{1/2}$~\cite{Lifshitz:1980} when both magnetic and
superconducting fluctuations are ignored. This simplest approximation
for the equilibrium value of $\psi$ is obtained from the GL free
energy~(\ref{eq1}) provided $e = 0$ and the gradient term is neglected.
Hereafter we shall keep the symbol $|\psi_0|$ for the equilibrium order
parameter in the more general case when the magnetic fluctuations are
not neglected and shall denote the same quantity for $e=0$ by $\eta
\equiv |\psi_0(e=0)| = (|a|/b)^{1/2}$.

The above described approximation neglects the saddle point solutions
of GL equations, where $\langle \psi(\vec{x}) \rangle$ is
$\vec{x}$-dependent. Therefore, the vortex state that is stable in type
II superconductors  cannot be achieved. This is consistent with the
choice of a zero external magnetic field, where the vortex state cannot
occur in any type superconductor. These arguments can be easily
verified with the help of GL equations~\cite{Lifshitz:1980} for a zero
external magnetic field; the only nonzero solution for $\psi$ in this
case is given by $\eta = (|a|/b)^{1/2}$ although the magnetic
fluctuations $\vec{A}(\vec{x}) = \delta \vec{A}(\vec{x})$ are properly
considered.

In conclusion we can argue that the described method will be convenient
for both type I and type II superconductors in a zero external magnetic
field, provided the $\psi$-fluctuations have a negligibly small effect
on phase transition properties $T_{c0}=T_{c}(H_0=0)$, where  $T_{c}$
denotes the phase transition line for any  $H_0 \ge 0$. For type II
superconductors in  $H_0 > 0$, two lines  $T_{c1}(H_0)$ and
$T_{c2}(H_0)$ should be defined, usually given by $H_{c1}(T)$ and
$H_{c2}(T)$~\cite{Lifshitz:1980}.

\subsection{Derivation of effective free energy}
\label{2.3.}

When the order parameter $\psi$ is uniform the functional~(\ref{eq1})
is reduced to
\begin{equation}
\label{eq11} F(\psi, \vec{A}) =  F_0(\psi) + F_A(\psi)\;,
\end{equation}
with
\begin{equation}
 \label{eq12}
  F_0(\psi) = V_D(a|\psi|^2 +\frac{b}{2}|\psi|^4)
 \end{equation}
and
\begin{equation}
\label{eq13} F_A(\psi) = \frac{1}{8\pi}\int d^Dx \left \{ \rho(\psi)
\vec{A}^2(\vec{x}) + \frac{1}{2} \sum_{i,j=1}^D \left( \frac{\partial
A_j}{\partial x_i} -\frac{\partial A_i}{\partial x_j} \right)^2
\right\} \;.
\end{equation}
Here $\rho = \rho_0 |\psi|^2$ and $\rho_0 = (8 \pi e^2/mc^2)$. It is
convenient to calculate the partition function $Z(\psi)$ and the
effective free energy ${\cal{F}}_{\mbox{\scriptsize D}}(\psi)$ in the
$\vec{k}$-space, where Eqs.~(\ref{eq9}) and~(\ref{eq13}) take the form
\begin{equation}
\label{eq14} \int_{- \infty}^{\infty} \prod_{j=1}^D
\prod_{\vec{k}>0}^{k \le \Lambda} d \mbox{Re} A_j(\vec{k}) d \mbox{Im}
A_j(\vec{k}) \delta \left[ \vec{k}\cdot \vec{A}(\vec{k}) \right ]
\end{equation}
and
\begin{equation}
\label{eq15} F_A(\psi) = F_A(0) + \Delta F_A(\psi)\;.
\end{equation}
Here
\begin{equation}
\label{eq16} F_A(0) = \frac{1}{8 \pi} \sum_{j,k} k^2 \left|A_j(\vec{k})
\right |^2 \;,
\end{equation}
and
\begin{equation}
\label{eq17} \Delta F_A(\psi)= \rho \sum_{j,k} \left|A_j(\vec{k})
\right |^2 \; ;
\end{equation}
note, that we have used the Coulomb gauge  $\vec{k}. \vec{A}(\vec{k}) =
0$.

Then the partition function~(\ref{eq8}) will be
\begin{equation}
\label{eq18} {\cal{Z}}(\psi) = e^{-F_0(\psi)/k_B T}{\cal{Z}}_A(\psi)\;,
\end{equation}
where
\begin{equation}
\label{eq19} {\cal{Z}}_A(\psi) = \int {\cal{D}}A e^{-F_A(\psi)/k_B T}
\end{equation}
with $F_A(\psi)$ given by~(\ref{eq17}) and the functional integration
is defined by the rule~(\ref{eq14}). With the help of Eqs.~(\ref{eq10})
- (\ref{eq19}) the effective free energy ${\cal{F}}_D(\psi)$ becomes
\begin{equation}
\label{eq20} {\cal{F}}_D(\psi) = F_0(\psi) + {\cal{F}}_f(\psi)\;,
\end{equation}
where $F_0(\psi)$ is given by Eq.~(\ref{eq12}) and
\begin{equation}
\label{eq21} {\cal{F}}_f(\psi) = -k_BT \ln{\left[
\frac{{\cal{Z}}(\psi)}{{\cal{Z}}(0)} \right]}
\end{equation}
is the $\psi$-dependent fluctuation part of  ${\cal{F}}(\psi)$. In
Eq.~(\ref{eq20}) the  $\psi$-independent fluctuation energy $\{-k_B T
\ln{\left[ {\cal{Z}}_A(0)\right]}\} $ has been omitted. This energy
should be ascribed to the normal state of the superconductor which, by
convention, is set equal to zero.

Defining the statistical averages
\begin{equation}
\label{eq22} \langle(...)\rangle = \frac{\int
{\cal{D}}{\cal{A}}e^{-F_A(0)/k_BT}(...)}{{\cal{Z}}_A(0)},
\end{equation}
we can write Eq.~(\ref{eq21}) in the form
\begin{equation}
 \label{eq23}
{\cal{F}}_f(\psi) = - k_BT \ln{\langle e^{- \Delta F_A(\psi)/k_BT }
\rangle}.
\end{equation}
Eq.~(\ref{eq23}) is a good starting point for the perturbation
calculation of ${\cal{F}}_f(\psi)$. We expand the exponent in
Eq.~(\ref{eq23}) and also take into account the effect of the logarithm
on the infinite series~\cite{Uzunov:1993} and obtain in result
\begin{equation}
 \label{eq24}
{\cal{F}}_f(\psi) = \sum_{l=1}^{\infty} \frac{(-1)^l}{l!
(k_BT)^{l-1}}\langle \Delta F_A^l(\psi) \rangle_c \;,
 \end{equation}
where $\langle...\rangle_c$ denotes connected
averages~\cite{Uzunov:1993}. Now we have to calculate averages of the
type
\begin{equation}
 \label{eq25}
\langle A_{\alpha}(\vec{k}_1),A_{\beta}(\vec{k}_2)...
A_{\gamma}(\vec{k}_n)\rangle_c \;.
\end{equation}
Here we shall use the Wick theorem and the correlation function of form
\begin{equation}
 \label{eq26}
G_{ij}^{(A)}(\vec{k}, \vec{k^{\prime}}) = \langle
A_i(\vec{k})A_j(-\vec{k^{\prime}}) \rangle =
\delta_{\vec{k},\vec{k^{\prime}}}G_{ij}^{A}(k)\;,
\end{equation}
where
\begin{equation}
 \label{eq27}
G_{ij}^{A}(k) = \langle A_i(k)A_j(-k) \rangle = \frac{4 \pi k_BT}{k^2}
\left( \delta_{ij} - \hat{k_i}  \hat{k_j}\right)
\end{equation}
 and $
\hat{k_i} = (k_i/k)$.

The calculation of lowest order terms $(l=1,2,3)$ in Eq.~(\ref{eq24})
with the help of ~(\ref{eq25})~-~(\ref{eq27}) is straightforward. The
infinite series~(\ref{eq24}) can be exactly summed up and the result is
the following logarithmic function
\begin{equation}
 \label{eq28}
{\cal{F}}_f(\psi) = \frac{(D-1)}{2}\: k_BT \sum_ k \ln
\left[1+\frac{\rho(\psi)}{k^2} \right]\;.
\end{equation}

The same result for ${\cal{F}}_f(\psi)$ can be obtained by a direct
calculation of the Gaussian functional integral~(\ref{eq9}). This is
done using the integral representation of $\delta$-function
in~(\ref{eq9}) or~(\ref{eq14}) but it introduces an additional
functional integration that should be carried out after the integration
over $A_j(\vec{x})$.

Eqs.~(\ref{eq10}),~(\ref{eq20}) and~(\ref{eq28}) give the
 effective free energy density
\begin{equation}
 \label{eq29}
f_D(\psi) = {\cal{F}}_D(\psi)/V_D
\end{equation}
in the form
\begin{equation}
 \label{eq30}
f_D(\psi) = f_0(\psi) + \Delta f_D(\psi)\;,
\end{equation}
where
\begin{equation}
 \label{eq31}
 f_0(\psi) = a|\psi|^2 + \frac{b}{2} |\psi|^4
\end{equation}
and
 \begin{equation}
 \label{eq32}
\Delta f_D(\psi) =  \frac{(D-1) k_BT}{2 V_D}\sum_ k \ln
\left(1+\frac{\rho}{k^2} \right)\;.
\end{equation}

Eqs.~(\ref{eq20}) and ~(\ref{eq29})~-~(\ref{eq32}) are the basis of our
further considerations. We should mention that the fluctuation
contribution $\Delta f_D(\psi)$  to $f(\psi)$ transforms to a
convergent integral in the continuum limit
\begin{equation}
\label{eq33} \frac{1}{V_D} \sum_k \to \int \frac{d^Dk}{(2 \pi)^D} = K_D
\int_0^{\Lambda} dk.k^{D-1} \;,
\end{equation}
where $K_D = 2^{1-D} \pi^{-D/2}/ \Gamma(D/2)$ for all spatial
dimensionalities $D \ge 2$. But the terms in the expansion of the
logarithm in~(\ref{eq32}) are power-type divergent with the exception
of several low-order terms in certain dimensionalities $D$. Therefore,
we shall work with a finite sum of an infinite series of infinite
terms. In our further calculations we shall keep the cutoff $\Lambda$
finite for all relevant terms in $\Delta f_D(\psi)$. This is the
condition to obtain correct results.

\subsection{Particular dimensions}
\label{sec.2.4.}

For purely 2D superconductor consisting of a single atomic layer, we
can use Eqs.~(\ref{eq29})-~(\ref{eq32}) setting $D=2$ and calculate
$\Delta f_2(\psi)$ with the help of the rule~(\ref{eq33}):
\begin{equation}
\label{eq34} \Delta f_2(\psi) = \left(\frac{k_BT}{8 \pi}\right)
\left[(\Lambda^2+ \rho_0|\psi|^2) \ln{\left
(1+\frac{\rho_0|\psi|^2}{\Lambda^2}\right)}
 -  \rho_0|\psi|^2 \ln{\left (\frac{\rho_0|\psi|^2}{\Lambda^2}\right)}
\right]\;.
\end{equation}
The first term of this free energy can be expanded in powers of
$|\psi|^2$:
\begin{equation}
\label{eq35} \Delta f_2(\psi) = \left(\frac{k_BT}{8
\pi}\right)\left\{\rho_0|\psi|^2+ \rho_0|\psi|^2 \ln{\left
(\frac{\Lambda^2}{\rho_0|\psi|^2}\right)}
 + \frac{\rho_0^2|\psi|^4}{2 \Lambda^2}\right \}\;.
\end{equation}

Thus we obtain the result from Ref.~\cite{Lovesey:1980}. This case is
of special interest because of the logarithmic term in the Landau
expansion for $f(\psi)$ but it has no practical application for the
lack of ordering in purely 2D superconductors.

For quasi-2D superconductors we assume that $(2\pi/\Lambda) > L_0 \gg
a_0$, where $L_0$ is the thickness of the superconducting film and a
more precise choice of the upper cutoff $\Lambda \ll (1/a_0)$ for the
wave numbers $k_i$ is a matter of an additional investigation
~\cite{Folk:2001} (see Sec.~2.1 and 2.5). In order to justify this
definition of a quasi-2D system we consider the more general case of a
3D system of volume $V = (L_1 L_2 L_0)$, where we can take the
continuum limit along the large dimensions ($L_1$ and $L_2$) of the
film because of the assumption $L_{\alpha} \gg (2\pi/\Lambda)$,
$(\alpha=1,2)$. The summation over the wave number $k_0= (2 \pi
n_0/L_0)$ cannot be substituted with an integration because $L_0 \ll
L_{\alpha}$ and the dimension $L_0$ does not obey the conditions, valid
for $L_{\alpha}$~\cite{Suzuki:1994,Suzuki:1995,Craco:1999}. Therefore,
for such 3D system we must sum over $k_0$ and integrate over two other
components ($k_1$ and $k_2$) of the wave vector $\vec{k}$. This gives
an opportunity for a systematic description of the 2D-3D crossover in
superconductors
~\cite{Shopova3:2003,Suzuki:1995,Craco:1999,Rahola:2001,Abreu:2003}
which fully justifies the application of more simple treatment for $a_0
\ll L_0 <(2\pi/\Lambda)$.

We consider the conditions $(2\pi/\Lambda) > L_0 \gg a_0$ as a
definition of a quasi-2D films of thickness $L_0$. The condition
$(2\pi/\Lambda) > L_0$  means that the sum in Eq.~(\ref{eq32}) contains
only terms with $(k_0 = 0)$. The summation over $\vec{k} = (k_1,k_2,0)$
gives a correct description of quasi-2D films of thickness $L_0$ and
this can be shown as a limiting case of the more general 2D-3D
crossover described in
Refs.~\cite{Shopova3:2003,Suzuki:1995,Craco:1999,Rahola:2001,Abreu:2003}.
Therefore, for a quasi-2D film we have the expression;
\begin{equation}
\label{eq36} \Delta f(\psi) = \frac{2}{L_0} \Delta f_2(\psi)\;,
\end{equation}
where  $\Delta f_2(\psi)$ is given by Eq.~(\ref{eq34}).

For the bulk (3D) superconductor we obtain:
\begin{equation}
\label{eq37} \Delta f_3(\psi) = \frac{k_BT}{2 \pi}\left[
\frac{\Lambda^3}{3}\ln{\left(1 + \frac{\rho_0|\psi|^2}{\Lambda^2}
\right)} +  \frac{2}{3}\rho_0|\psi|^2 \Lambda - \frac{2}{3}\rho_0^{3/2}
|\psi|^3 \arctan{\left( \frac{\Lambda}{\sqrt{\rho_0 |\psi|^2}}\right)}
\right].
\end{equation}
For the Landau expansion in powers of $|\psi|$ this form of $f_3(\psi)$
confirms the respective results in Refs.~\cite{Halperin:1974,Chen:1978}
and moreover correctly gives a term of type $\rho^2_0 |\psi|^4$ which
was supposed small and neglected in these preceding papers. This
problem will be discussed in Sec.~3.

For 4D-systems $\Delta f_{\mbox{\scriptsize D}}(\psi)$ becomes
\begin{equation}
\label{eq39} \Delta f_4(\psi) = \frac{3k_BT}{64 \pi^2}\left[\Lambda^2
\rho_0 |\psi|^2 + \Lambda^4 \ln{\left(1 +
\frac{\rho_0|\psi|^2}{\Lambda^2} \right)} - \rho_0^2|\psi|^4
\ln{\left(1 +
 \frac{\Lambda^2}{\rho_0|\psi|^2} \right)}
\right].
\end{equation}
The above expression for $ \Delta f_4(\psi)$ can be also expanded in
powers of $|\psi|$ to show that it contains a term of the type
$|\psi|^4 \ln{(\sqrt{\rho_0}|\psi|/\Lambda)}$ which produces a first
order phase transition; this case is considered in  the scalar
electrodynamics~\cite{Coleman:1973}. In our further investigation we
shall focus our attention on 3D and quasi-2D superconductors.

The free energy density $\Delta f_{\mbox{\scriptsize D}}(\psi)$ can be
expanded in powers of $|\psi|$ but the Landau expansion can be done
only in an incomplete way for even spatial dimensions. Thus
$f_2(\psi)$, $f_4(\psi)$, and $f(\psi)$ - the free energy density
corresponding to the quasi-2D films, contain logarithmic terms which
should be kept in their original form in the further treatment of the
function $\Delta f_{\mbox{\scriptsize D}}(\psi)$ in the Landau
expansion. We shall do our analysis in two ways: with and without
Landau expansion of $\Delta f_D (\psi)$. These variants of the theory
will be called ``exact" theory (ET) and ``Landau" theory (LT),
respectively. We shall show that these two ways of investigation give
the same results in all cases except for quasi-2D films with relatively
small thicknesses ($L_0 \ll \xi_0$). It seems important to establish
the differences between two variants of the theory because the HLM
effect is very small and any incorrectness in the theoretical analysis
may be a cause for an incorrect result. By same arguments we  shall
investigate the effect of the factor $T$ in $\Delta
f_{\mbox{\scriptsize D}}(\psi)$ on the thermodynamics of  quasi-2D
films. This factor can be represented as $T= T_{c0}(1 + t_0)$ and one
may expect that the usual approximation $T\approx T_{c0}$, which is
well justified  in the Landau theory of phase
transitions~\cite{Lifshitz:1980,Uzunov:1993}, may be applied. We shall
show for both 3D and quasi-2D superconductors, that this way of
approximation can be made by neglecting terms in the thermodynamic
quantities smaller than the leading ones. On the other hand practical
calculations lead to the conclusion that this approximation cannot be
made without a preliminary examination because for some quasi-2D films
it produces a substantial error of about 10$\%$. LT, in which the
factor $T$ is substituted by $T_{c0}$, will be called a ``simplified
Landau expansion" - SLT.

\subsection{Validity}
\label{sec.2.45.}

The general result~(\ref{eq29})~-~(\ref{eq32}) for the effective free
energy $f(\psi)$ has the same domain of validity~\cite{Lifshitz:1980}
as the GL free energy functional in a zero external magnetic field.
When we neglect a sub-nano interval of temperatures near the phase
transition point we can use Eq.~(\ref{eq1}) provided $|t_0|=|T -
T_{c0}|/T_{c0} <1$, or in the particular case of type I
superconductors, $|t_0|< \kappa^2$~\cite{Lifshitz:1980}. Note, that the
latter inequality does not appear in the general GL approach. It comes
as a condition for the consistency of this approach with the
microscopic BCS theory for type I superconductors~\cite{Lifshitz:1980}.

Taking the continuum limit we have to assume that all dimensions of the
body, including the thickness $L_0$, are much larger than the
characteristic lengths $\xi$ and $\lambda$. The exception of this rule
is when we consider thin films. Especially for thin films of type I
superconductors, where ($(2\pi/\Lambda)
> L_0 \gg a_0$),  we should have in mind that
$\xi(T) >\lambda(T)$, so the inequalities $\xi> \lambda
> \xi_0 > \lambda_0$ hold true in the domain of validity of the GL theory
 $|t_0| < \kappa^2 < 1$.
In Ref.~\cite{Folk:2001} a comprehensive choice of the cutoff $\Lambda$
has been made ($\Lambda = \xi_0$) and we shall discuss this point in
Sec.~3 and 4. Note, that the respective conditions for quasi-2D films
of type II superconductors are much weaker and are reduced to the usual
requirements: $\kappa > 1/\sqrt{2}$, $|t_0| < 1$ and $(2\pi/\Lambda) >
L_0 \gg a_0$.

 If we do a Landau expansion of $f_D(\psi)$, in powers of
$|\psi|^2$ the condition $\rho \ll \Lambda^2$ should be satisfied. In
order to evaluate this condition we substitute $|\psi|^2$ in $\rho =
\rho_0|\psi|^2$ with $\eta^2 =|a|/b$ which corresponds to $e=0$
(Sec.~2.2). As $\lambda^2(T)=1/\rho$, the condition for the validity of
the Landau expansion becomes $[\Lambda \lambda(T)]^2 \gg 1$, i. e.,
$(\Lambda \lambda_0)^2 \gg |t_0|$. Choosing the general form of
$\Lambda_{\tau} = (\pi \tau/\xi_0)$ where $\tau$ describes the
deviation of $\Lambda_{\tau}$ from $\Lambda_1 \equiv \Lambda = (\pi
/\xi_0)$, we obtain $(\pi \tau \kappa)^2 \gg |t_0|$ ; $\kappa =
(\lambda_0 /\xi_0)$ is the GL parameter.

Thus we can conclude that in type II superconductors, where~$\kappa =
(\lambda_0/\xi_0)> 1/\sqrt{2}$,~the condition $(\rho/\Lambda^2) \ll
1$~is satisfied very well for values of the cutoff in the interval
between $\Lambda = (\pi/\xi_0)$ and $\Lambda = (\pi/\lambda_0)$, i.e.,
for $1<\tau<(1/\kappa)$. For type I superconductors, where $\kappa <
1/\sqrt{2}$ the cutoff values $\Lambda \sim (1/\xi_0)$ leads to the
`BCS condition ($|t_0| < \kappa^2$) for the validity of the GL
approach. Substantially larger cutoffs ($\Lambda \gg \pi/\xi_0$), for
example, $\Lambda \sim (1/\lambda_0)$ for type I superconductors with
$\kappa \ll 1$ lead to a contradiction of this BCS condition with the
requirement $\rho \ll \Lambda^2$. This inconsistency will be discussed
again in Sec.~3.3.

 In our calculations we often
use another parameter $ \mu_{\tau} = (1/\pi \tau \kappa)^2$ and, in
particular,  $\mu \equiv \mu_1 = (1/\pi \kappa)^2$ and in terms of
$\mu$ the condition for the validity of expansion of $f_D(\psi)$
becomes $\mu |t_0| \ll 1$, or, more generally, $\mu_{\tau}|t_0| \ll 1$.
Choosing $\tau = 1/\pi$ we obtain the BCS criterion for the validity of
the GL free energy of type I superconductors~\cite{Lifshitz:1980}. The
choice $\tau = (\xi_0/\pi\lambda_0)$ corresponds to the cutoff
$\Lambda_{\tau} = 1/\lambda_0$. As we shall see in Sec.~3 and~4 the
thermodynamics near the phase transition point has no substantial
dependence on the value of the cutoff $\Lambda_{\tau}$ but it should be
chosen in a way that is consistent with the mean-field-like
approximation.

Alternatively, the inequality $(\rho/\Lambda^2) \ll 1$ may be
investigated with the help of the reduced order parameter $\varphi$
defined by $\varphi = |\psi|/\eta_0$, where $\eta_0 \equiv \eta(T=0)=
(\alpha_0T_{c0}/b)^{1/2}$ is the so-called zero-temperature value of
the order parameter within the GL free energy $f_0(\psi)$, given by
Eq.~(\ref{eq31}); see also Sec.~2.2. The reduced order parameter
$\varphi$ will be equal to $|t_0|$ for $t_0 <0$, if only the magnetic
fluctuations are ignored, i.e., when $|\psi| = \eta$. Using the
notation $\varphi$, we obtain the condition $(\rho/\Lambda^2) \ll 1$ in
the form $\mu_{\tau}\varphi^2 \ll 1$. This condition seems to be more
precise because it takes into account the effect of magnetic
fluctuations on the order parameter $\psi$.

\section{BULK SUPERCONDUCTORS}
\label{sec.3.}

\subsection{Free energy}
\label{sec.3.1.}

The effective free energy  $f_3(\psi)$ of bulk (3D-) superconductors is
given by Eqs.~(\ref{eq29})~-~(\ref{eq31}) and~(\ref{eq37}). The
analytical treatment of this free energy can be done by Landau
expansion in small $(\sqrt{\rho_0}|\psi|/\Lambda)$. Up to order
$|\psi|^6$ we obtain
\begin{equation}
\label{eq39} f_3(\psi) \approx a_3 |\psi|^2 + \frac{b_3}{2} |\psi|^4 -
q_3|\psi|^3 + \frac{c_3}{2} |\psi|^6 \;,
\end{equation}
where
\begin{equation}
\label{eq40}
 a_3 = a + \frac{k_BT \Lambda \rho_0}{2\pi^2}\;,
\end{equation}
\begin{equation}
\label{eq41} b_3 = b + \frac{k_BT  \rho_0^2}{2\pi^2 \Lambda}\;,
\end{equation}
\begin{equation}
\label{eq42} q_3 = \frac{k_BT  \rho_0^{3/2}}{6 \pi}\;,
\end{equation}
and
\begin{equation}
\label{eq43} c_3 = - \frac{k_BT  \rho_0^{3}}{6 \pi^2 \Lambda^3}\;.
\end{equation}
The cutoff $\Lambda$ in Eqs.~(\ref{eq40}) - (\ref{eq43}) is not
specified and can be written in the form $\Lambda_{\tau} =
(\pi\tau/\xi_0)$ as suggested in Sec.~2.5.

 We shall just outline the
analysis of the above free energy. It can be shown by both analytical
and numerical calculations~\cite{Shopova:2002} that $|\psi|^6$-term has
no substantial effect on the thermodynamics, described by the free
energy~(\ref{eq39}). That is why we ignore this term and do the
analysis in the standard way~\cite{Uzunov:1993}. The possible phases
$|\psi_0|$ are found  as a solution of the equation of state:
\begin{equation}
\label{eq44} \left[ \partial f(\psi) / \partial |\psi| \:\right
]_{\psi_0} = 0\;.
\end{equation}
There always exists a normal phase $|\psi_0| = 0$ which gives a minimum
of  $f_3(\psi)$ for $a_3 >0$. The possible superconducting phases are
given by
\begin{equation}
\label{eq45} |\psi_0|_{\pm} = \frac{3q_3}{4b_3} \left( 1 \pm
\sqrt{1-\frac{16a_3b_3}{9q_3^2}}\right) \ge 0.
\end{equation}
Having in mind  the existence and stability conditions of
$|\psi_0|_{\pm}$-phases ~\cite{Uzunov:1993}, we obtain that the
$|\psi_0|_{+}$-phase exists for $(16a_3b_3) \le 9q_3^2$ and this region
of existence always corresponds to a minimum of   $f_3(\psi)$. The
$|\psi_0|_{-}$-phase exists for $0 < a_3 < (9q_3^2/16b_3)$ and this
region of existence always corresponds to a maximum of   $f_3(\psi)$,
i.e., this phase is absolutely unstable. For $a_3 =0,\; |\psi_0|_{-}=0$
and hence, coincides with the normal phase. For $9q_3^2 = (16a_3b_3)$
we have $|\psi_0|_{+} = |\psi_0|_{-}=(3q_3/4b_3)$  and $
f_3(|\psi_0|_{+}= f_3|\psi_0|_{-})= (27q_3^4/512b_3^3)$. Furthermore
$f_3(|\psi_0|_{-}) > 0$ for all allowed values of   $|\psi_0|_{-} > 0$,
whereas
\begin{equation*}
f_3(|\psi_0|_{+}) < 0  \;\;\; \mbox{for} \;\;\; a_3<(q_3^2/2b_3)\;,
\end{equation*}
and
\begin{equation*}
f_3(|\psi_0|_{+}) > 0  \;\;\; \mbox{for} \;\;\;(q_3^2/2b_3) <  a_3<
\frac{9q_3^2}{16b_3}\;.
\end{equation*}
The equilibrium temperature $T_{\mbox{\scriptsize eq}}$ of the first
order phase transition is defined by the equation $f(|\psi_0|_{+})= 0$
which gives the following result:
\begin{equation}
\label{eq46}
 2b_3(T_{\mbox{\scriptsize eq}})a_3(T_{\mbox{\scriptsize eq}}) =
  q_3^2(T_{\mbox{\scriptsize eq}})\;.
 \end{equation}
These results are confirmed by numerical calculations of the effective
free energy~(\ref{eq39})~\cite{Shopova:2002}; there also the influence
of the $|\psi|^6$-term is evaluated.

\subsection{Entropy and specific heat capacity}
\label{sec.3.3.}

The equilibrium entropy jump is $\Delta S = V \Delta s$ and $\Delta s =
- (df_3(|\psi|)/dT)$ can be calculated with the help of
Eq.~(\ref{eq39}) and the equation of state~(\ref{eq44}):
\begin{equation}
\label{eq47} \Delta s = - |\psi_0|^2 \Phi(|\psi_0|) \;,
\end{equation}
where $\Phi(|\psi_0|)$ is the following function:
\begin{equation}
\label{eq48}
\Phi(y) = (\alpha_0+\frac{k_B \Lambda \rho_0}{2 \pi^2}) -
\frac{\rho_0^{3/2}k_B}{6 \pi} y + \left(\frac{k_B \rho_0^2}{4 \pi^2
\Lambda}\right)  y^2\;.
\end{equation}

The specific heat capacity per unit volume $\Delta C = T(\partial\Delta
s/ \partial T)$ is obtained from~(\ref{eq47})
\begin{equation}
\label{eq49} \Delta C = - \left( \frac{T}{T_{c0}}\right )
\frac{\partial |\psi_0|^2}{ \partial t_0} \Phi(|\psi_0|) \;.
 \end{equation}
The quantities  $\Delta s(T)$ and $\Delta C(T)$ can be evaluated  at
the equilibrium phase transition point $T_{\mbox{\scriptsize eq}}$
which is found from Eq.~(\ref{eq46}):
\begin{equation}
\label{eq50} \frac{T_{\mbox{\scriptsize eq}}}{T_{c0}} \approx 1-
\frac{k_B \rho_0 \Lambda}{2 \pi^2 \alpha_0} + \frac{\left(
\rho_0^{3/2}k_B/6 \pi \right)^2 }{b+\left(\rho_0^{2}k_B/2 \pi^2 \Lambda
\right)T_{c0}} \left( \frac{T_{c0}}{\alpha_0}\right)\;,
  \end{equation}
provided $|\Delta T_c| =|T_{c0}-T_{\mbox{\scriptsize eq}}| \ll T_{c0}$.
Further we shall see  that the condition $|\Delta T_c| \ll T_{c0}$ is
valid in real substances. The second term in  r.h.s. of
Eq.~(\ref{eq50}) is a typical negative fluctuation contribution whereas
the positive third term in  r.h.s. of the same equality is typical for
first-order transitions~\cite{Uzunov:1993}.

To obtain the jumps  $\Delta s$ and $\Delta C$ at $T_{\mbox{\scriptsize
eq}}$ we have to put the solution $|\psi_0|_{+}$ found from
Eq.~(\ref{eq45}) in Eqs.~(\ref{eq47})~-~(\ref{eq49}). The result will
be:
\begin{equation}
\label{eq51} \Delta s = - \frac{q_{3c}^2}{b_{3c}^2}\left \{ \alpha_0 +
\frac{k_B\rho_0 \Lambda}{2 \pi^2}- \left( \frac{k_B\rho_0^{3/2} }{6
\pi}\right)^2 \frac{T_{\mbox{\scriptsize eq}}}{b_{3c}^2} \right \}\;,
\end{equation}
 and
\begin{equation}
\label{eq52} \Delta C = \frac{4 \alpha_0}{b_{3c}}\left(\alpha_0 T_{c0}
- \frac{q_{3c}^2b}{b_{3c}^2}\right ) \;,
 \end{equation}
 where $b_{3c}$ and $q_{3c}$ are the parameters $b_{3}$ and $q_{3}$  at
 $T=T_{\mbox{\scriptsize eq}}$. As $|\Delta T_c| =|T_{c0}-T_{\mbox{\scriptsize eq}}|
  \ll T_{c0}$ we can set
 $T_{\mbox{\scriptsize eq}} \approx T_{c0}$ in r.h.s. of
 Eqs.~(\ref{eq51})~and~(\ref{eq52}) and
  obtain $q_{3c} \equiv q_3(T = T_{\mbox{\scriptsize eq}})
 \approx q_3(T_{c0})$ and $b_{3c} \approx b_3(T_{c0})$.

 The latent heat $Q=- VT_{\mbox{\scriptsize eq}} \Delta s$ of the
 first order phase transition at $T_{\mbox{\scriptsize eq}}$ can be calculated from
 Eq.~(\ref{eq51}).
 If we neglect the charge $(e=0)$ which means to set $\rho_0=q_3=0$ and
 $b_{c3}=b$ in Eqs.~(\ref{eq51})~-~(\ref{eq52}) we shall get the
 result from Ref.~\cite{Halperin:1974} for the ratio
 \begin{equation}
 \label{eq53}
 (\Delta T)_{\mbox{\scriptsize eq}} =\frac{Q}{T_{\mbox{\scriptsize eq}} \Delta C} \;.
 \end{equation}
 Here we should mention that Eq.~(\ref{eq52}) gives the jump $\Delta C$
 at the equilibrium phase transition point of the first order phase
 transition, described by $|\psi|^3$ term~\cite{Uzunov:1993}, while $\Delta C$
 calculated in~Ref.~\cite{Halperin:1974} is equal to the specific heat
 jump at the standard second order transition $\Delta C = (\alpha_0^2T_{c0}/b)$
 and is four times smaller. Therefore, we obtain  $(\Delta T)_{\mbox{\scriptsize eq}}$ four
 times smaller than the respective value in Ref.~\cite{Halperin:1974}.

\subsection{Numerical values for Al}

\label{sec.3.3} In order to do the numerical estimates we represent the
Landau parameters $\alpha_0$ and $b$  with the help of the
zero-temperature coherence length $\xi_0$ and the zero-temperature
critical magnetic field $H_{c0}$. The connection between them is given
by formulae of the standard GL theory of
superconductivity~\cite{Lifshitz:1980}: $\xi_0^2=(\hbar^2/4m\alpha_0
T_{c0})$ and $H_{c0}^2=(4 \pi \alpha_0^2 T_{c0}^2/b)$. The expression
for the zero-temperature penetration depth $\lambda_0=(\hbar c/2
\sqrt{2} e H_{c0}\xi_0 )$ is obtained from the above relation and
$\lambda_0=(b/\alpha_0 T_{c0} \rho_0)^{1/2}$. We shall use the
following experimental values of  $T_{c0}$, $H_{c0}$ and $\xi_0 $ for
Al: $ T_{c0}=1.19$K, $H_{c0}=99$Oe, $\xi_0=1.6 \mu$m, $\kappa=0.01$
~\cite{Halperin:1974, Madelung:1990}. The experimental values for
$T_{c0}$, $H_{c0}$ and $\xi_0$ vary about 10-15$\%$ depending on the
method of measurement and the geometry of the samples (bulk material or
films) but such deviations do not affect the results of our numerical
investigations.

The evaluation of the parameters $a_3$ and $b_3$ for Al
gives:
\begin{equation}
\label{eq54} a_3 = (\alpha_0 T_{c0})\left[t_0 + 0.972 \times 10^{-4}
(1+t_0)\tau\right],
\end{equation}
and
\begin{equation}
\label{eq55}
\frac{b_3}{b} =1 +\frac{0.117}{\tau}\;.
\end{equation}
Setting $\tau=1$ corresponds to the cutoff $\Lambda_1 =(\pi/\xi_0)$
(Sec~2.5). For $\tau =(1/\kappa)_{Al} = 10^2$ which corresponds to the
much higher cutoff $\Lambda =(\pi/\lambda_0)$ we have $b_3 \approx b$.
i.e., the~$\rho_0^2$~-term in $b_3$, given by Eq.~(\ref{eq41}), can be
neglected. However, as we see from Eq.~(\ref{eq55}), for $\tau = 1$ the
same $\rho_0^2$-correction in the parameter $b_3$ is of order $0.1b$
and cannot be automatically ignored in all calculations, in contrast to
the supposition in Refs.~\cite{Halperin:1974, Chen:1978}. However, the
more important fluctuation contribution in 3D superconductors comes
from the $\tau-$term in Eq.~(\ref{eq54}) for the parameter $a_3$. This
term is of order $10^{-4}$ for $\tau \sim 1$ and this is consistent
with the condition $|t_0| < \kappa^2 \sim 10^{-4}$ but for $\tau \sim
10^2$, i.e., for $\Lambda \sim (\pi/\lambda_0) \sim 10^6\mu$m, the same
$\tau-$ term is of order $10^2$ which exceeds the temperature interval
($T_{c0} \pm 10^{-4}$) for the validity of  BCS condition of Al
(Sec.~2.5).

These results demonstrate that for our theory to be consistent, we must
choose the cutoff $\Lambda_{\tau}=(\pi \tau/\xi_0)$, where  $\tau$ is
not a large number $(\tau \to 1 \div 10)$. To be more concrete we set
$\Lambda=\Lambda_1=(\pi/\xi_0)$ as suggested in Ref.~\cite{Folk:2001}.

The temperature shift $t_{\mbox{\scriptsize eq}} =
t_0(T_{\mbox{\scriptsize eq}})$ for bulk Al can be estimated with the
help of Eq.~(\ref{eq50}). We obtain that this shift is negative and
very small: $t_{\mbox{\scriptsize eq}} \sim - 10^{-4}$. Note, that the
second term in the r.h.s. of Eq.~(\ref{eq50}) is of order $10^{-4}$
provided $\Lambda \sim (1/\xi_0)$ whereas the third term in the r.h.s.
of the same equality is of order $10^{-5}$. Once again the change of
the cutoff $\Lambda$ to values much higher than ($\pi/\xi_0$) will take
the system outside the temperature interval where the BCS condition for
Al is valid. Let us note, that in Ref.~\cite{Shopova:2002} the
parameter $t$ corresponds to our present notation $t_0$. But the
numerical calculation of the free energy function $f_3(\psi)$ in
Ref.~\cite{Shopova:2002} was made for the SLT variant of the theory and
the shifted parameter  ($t_0 + 0.972\times 10^{-4}$) was incorrectly
identified with  $t$ and this lead to the wrong conclusion for its
positiveness at the equilibrium phase transition point
$T_{\mbox{\scriptsize eq}}$. As a matter of fact, the shifted parameter
($t_0 + 0.972\times 10^{-4}$) is positive at $T_{\mbox{\scriptsize
eq}}$ but $t_{\mbox{\scriptsize eq}} \equiv t_0(T_{\mbox{\scriptsize
eq}})$ is negative.

Having in mind these remarks, when we evaluate $\Delta  s$ and  $\Delta
C$ for bulk Al we can use simplified versions of ~(\ref{eq51}) and
~(\ref{eq52}) which means to consider only the first terms in the r.h.s
and  to take $q_{3c} \approx q_3$   and $b_{3c} \approx b$ at $T_{c0}$.
In this way we obtain
\begin{equation}
\label{eq56} Q = - T_{c0}\Delta s  = 0.8 \times 10^{-2}
\left[\frac{\mbox{erg}}{\mbox{K}\:.\: \mbox{cm}^3}\right]\;,
\end{equation}
and
\begin{equation}
\label{eq57} \Delta C  = 2.62 \times 10^{3}
\left[\frac{\mbox{erg}}{\mbox{cm}^3}\right]\;.
\end{equation}
The results are consistent with an evaluation of $\Delta C$ for Al as a
jump ($\Delta \tilde{C} = \alpha_0^2T_{c0}/b$) at the second order
superconducting transition point~\cite{Halperin:1974} that, as we
mentioned above, is four times smaller than the jump  $\Delta C$ given
by Eq.~(\ref{eq57}).

A complete numerical evaluation of the function $f_3(\psi)$ and the
jump of the order parameter at $T_{\mbox{\scriptsize eq}}$ for bulk Al
was presented for the first time in Ref.~\cite{Shopova:2002}. The
results there confirm that the order parameter jump and Q for bulk type
I superconductors are very small and can hardly be observed in
experiments.

We shall finish the presentation of bulk Al with a discussion of the
ratio~(\ref{eq53}). It can be also written in the form
\begin{equation}
\label{eq58}
 (\Delta T)_{\mbox{\scriptsize eq}} = \frac{32\pi}{9}  \left(\frac{T_{c0}^2}{
b \alpha_0}\right) \left(\frac{e^2}{ m c^2}\right)^3,
\end{equation}
and it differs by a factor $1/4$ from the respective result in
Ref.~\cite{Halperin:1974}. This difference is due to the fact that we
take $\Delta C$ as the jump at the first order transition temperature
$T_{\mbox{\scriptsize eq}}$ while in the above cited
paper~\cite{Halperin:1974} the authors define $\Delta C$ as a
hypothetic jump ($\Delta\tilde{C}$) at the standard second order phase
transition point. From Eq.~(\ref{eq56}) we obtain
\begin{equation}
\label{eq59} (\Delta T)_{\mbox{\scriptsize eq}} = 6.7 \times 10^{-12}
(T_c^3 H_{c0}^2 \xi_0^6),
\end{equation}
and multiplying the number coefficient in the above expression by 4 we
can obtain Eq.~(10) from Ref.~\cite{Halperin:1974}.

\section{QUASI-2D FILMS}
\label{sec.4.}

Thin quasi-2D films $(a_0 \ll L_0 < 2\pi/\Lambda)$ can be investigated
with the help of the respective free energy density $ f(\psi)$ given by
Eqs.~(\ref{eq30}) and~(\ref{eq31}), $\Delta f_D(\psi)$ is taken from
Eqs.~(\ref{eq34}) and~(\ref{eq36}). The free energy of quasi-2D
superconducting films was derived and analyzed for the first time in
Ref.~\cite{Folk:2001} using the Landau expansion of  $\Delta f_2(\psi)$
in powers of $|\psi|^2$; see Eq.~(\ref{eq35}). As is shown for the
first time by Lovesey~\cite{Lovesey:1980} in the simple 2D case the
fluctuation contribution $\Delta f_{\mbox{\scriptsize D}}(\psi)$, of
form given by Eq.~(\ref{eq35}), leads to a fluctuation-induced first
order phase transition. In contrast to 3D superconductors where the
first order of the phase transition is generated  by  $|\psi|^3$-term
in $\Delta f_3(\psi)$, in 2D superconductors the first order of the
phase transition is a result of the presence of $|\psi|^2
\mbox{ln}|\psi|$ in Eq.~(\ref{eq35}). But the Meissner phase cannot
occur in 2D (single atomic layer) superconductors because of the strong
fluctuations and hence this case is of no interest. In quasi-2D films,
where the Meissner phase does occur for properly chosen thickness  of
the film $(L_0 \ll 2\pi/\Lambda)$~\cite{Folk:2001}, the  change of the
order of normal-to-superconducting phase transition is better
pronounced than in bulk superconductors. This is well illustrated in
the above cited paper~\cite{Folk:2001} by numerical data for Al films
with thickness $L_0=0.1 \mu \mbox{m}$. Following
Refs.~\cite{Folk:2001,Shopova1:2003} and the arguments presented in
Sec.~3.3 we shall choose the cutoff $\Lambda = \pi/\xi_0$.

The expansion of the respective free energy in powers of $|\psi|$ leads
to somewhat clumsy analysis and for this reason we shall use the
approach in Ref.~\cite{Shopova1:2003,Shopova2:2003} where the quasi-2D
films have been investigated with the help of the general form of
$\Delta f(\psi)$ described by Eq.~(\ref{eq34}). In both variants of the
theory (ET and LT; see Sec.~2.4) the thickness $L_0$ of the quasi-2D
film has an effect on the thermodynamic behavior, that is similar to
the influence of material parameters  $\alpha_0$ and $b$. This is very
well seen in the Landau expansion~(\ref{eq35}) of the free energy
$f(\psi)$ given by (\ref{eq36}), where the parameters $a$ and $b$
acquire a fluctuation contribution that depends on $L_0$. The influence
of $L_0$ on the thermodynamic properties can be considered as a
characteristic feature of quasi-2D systems~\cite{Craco:1999,
Rahola:2001}, a feature, absent in purely 2D films \cite{Lovesey:1980}.
It is unambiguously demonstrated by several theoretical studies of the
2D-3D crossover in systems with slab geometry ~\cite{Shopova3:2003,
Craco:1999, Rahola:2001, Abreu:2003} that the $L_0$--dependence as
given in Eq.~(\ref{eq36}) correctly describes quasi-2D films.

Following Refs.~\cite{Shopova1:2003, Shopova2:2003} and having in mind
the above discussion we can present the free energy density
$f(\psi)=(F(\psi)/L_1L_2)$ in the form
\begin{equation}\label{eq60}
 f(\varphi) =  \frac{H_{c0}^2}{8 \pi}\left[2 t_0 \varphi^2 + \varphi^4
 + C(1+ t_0 ) \Gamma(\mu \varphi^2) \right],
\end{equation}
where
\begin{equation}\label{eq61}
  \Gamma(y) = (1+y) \ln{(1+y)} -y \ln{y},
\end{equation}

\begin{equation}\label{eq62}
  C=\left(\frac{2 \pi^2 k_B T_{c0}}{L_0 \xi_0^2 H_{c0}^2}\right).
\end{equation}

In Eqs.~(\ref{eq60})~-~(\ref{eq62}) we have set $\Lambda=(\pi/\xi_0)$
and introduced the notation $\varphi = |\psi|/\eta_0$; the quantity
$\eta_0$ is defined in Sec.~2.5. Some of the properties of free
energy~(\ref{eq60}) were analyzed in Ref.~\cite{Shopova1:2003} for Al
films and in Ref.~\cite{Shopova2:2003} for films of Tungsten (W),
Indium (In), and Aluminium (Al). Here we shall summarize and justify
the preceding results and, moreover, we shall present new results about
the properties of the Landau expansion of effective free energy. Note,
that the function $\Gamma(y)$ cannot be fully expanded in powers of $y$
because of the term of type ($y\:\mbox{ln}y$) in Eq.~(\ref{eq61}).

The extensive investigations~\cite{Shopova1:2003,Shopova2:2003} of
films of W, Al and In with thicknesses from 0.05 $\mu$m to 2 $\mu$m
confirm the intuitive notion that the HLM effect is stronger for
smaller values of $L_0$. The numerical analysis shows that type I
superconductors with relatively small GL parameter $\kappa$ and
relatively high critical field $H_{c0}$ may be the best candidates for
the experimental observation of the effect. The best material from the
above enumerated substances seems to be Al; tungsten has an extremely
small GL parameter but also a small critical field that makes it
inconvenient for experiments. The relatively high $H_{c0}$ of In
results in relatively large latent heat, $Q \sim 4.0\;
(\mbox{erg}/\mbox{cm}^3)$ but for films with $L_0 \sim 0.05\; \mu
\mbox{m}$ the order parameter jump $|\psi|_{\mbox{\scriptsize eq}} =
\varphi_{\mbox{\scriptsize eq}}\eta_0$ at $T_{\mbox{\scriptsize eq}}$
is twice smaller than that for the respective Al
films~\cite{Shopova2:2003}: $|\psi|_{\mbox{\scriptsize eq}} =
0.05\times 10^{-11}$ for In and $|\psi|_{\mbox{\scriptsize eq}} =
0.1\times 10^{-11}$ for Al. We have to stress the role of critical
magnetic field $H_{c0}$, a fact established for the first time
in~\cite{Shopova2:2003} and the present paper.

With the help of data from Refs.~\cite{Folk:2001, Shopova:2002,
Shopova1:2003, Shopova2:2003} we compared the thermodynamic quantities
near the first order phase transition point in bulk Al and Al films of
$L_0 \sim 0.1 \mu \mbox{m}$. They are given in Table 1, where
$t_{\mbox{\scriptsize eq}} =t_0(T_{\mbox{\scriptsize eq}})$,
$\varphi_{eq} =\varphi (T_{\mbox{\scriptsize eq}})$ is the equilibrium
jump of the reduced order parameter and $|\psi|_{\mbox{\scriptsize eq}}
=\varphi_{\mbox{\scriptsize eq}}\eta_0$ is the order parameter jump at
$T_{\mbox{\scriptsize eq}}$.

\vspace{0.5cm}
 \small
 Table 1. Numerical data for bulk Al and Al film of thickness $L_0=0.1\mu$m\\
 \vspace{0.5cm}
\begin{tabular}{lcccl}
\hline quantity& $t_{\mbox{eq}}$  &$\varphi_{\mbox{eq}}$ &
$|\psi|_{\mbox{eq}}$
 & Q $(\mbox{erg}/\mbox{cm}^3)$  \\ \hline  bulk Al
& $-0.492 \times 10^{-4}$
& $0.0032$ & $0.8 \times 10^9 $& $ 0.8 \times 10^{-2}$  \\
\hline $L_0=0.1\mu$m& $-0.00147$ & $0.032$ & $0.8  \times 10^{10}$ &
$0.8$ \\
\hline
\end{tabular}

\newpage
\begin{figure}
\begin{center}
\epsfig{file=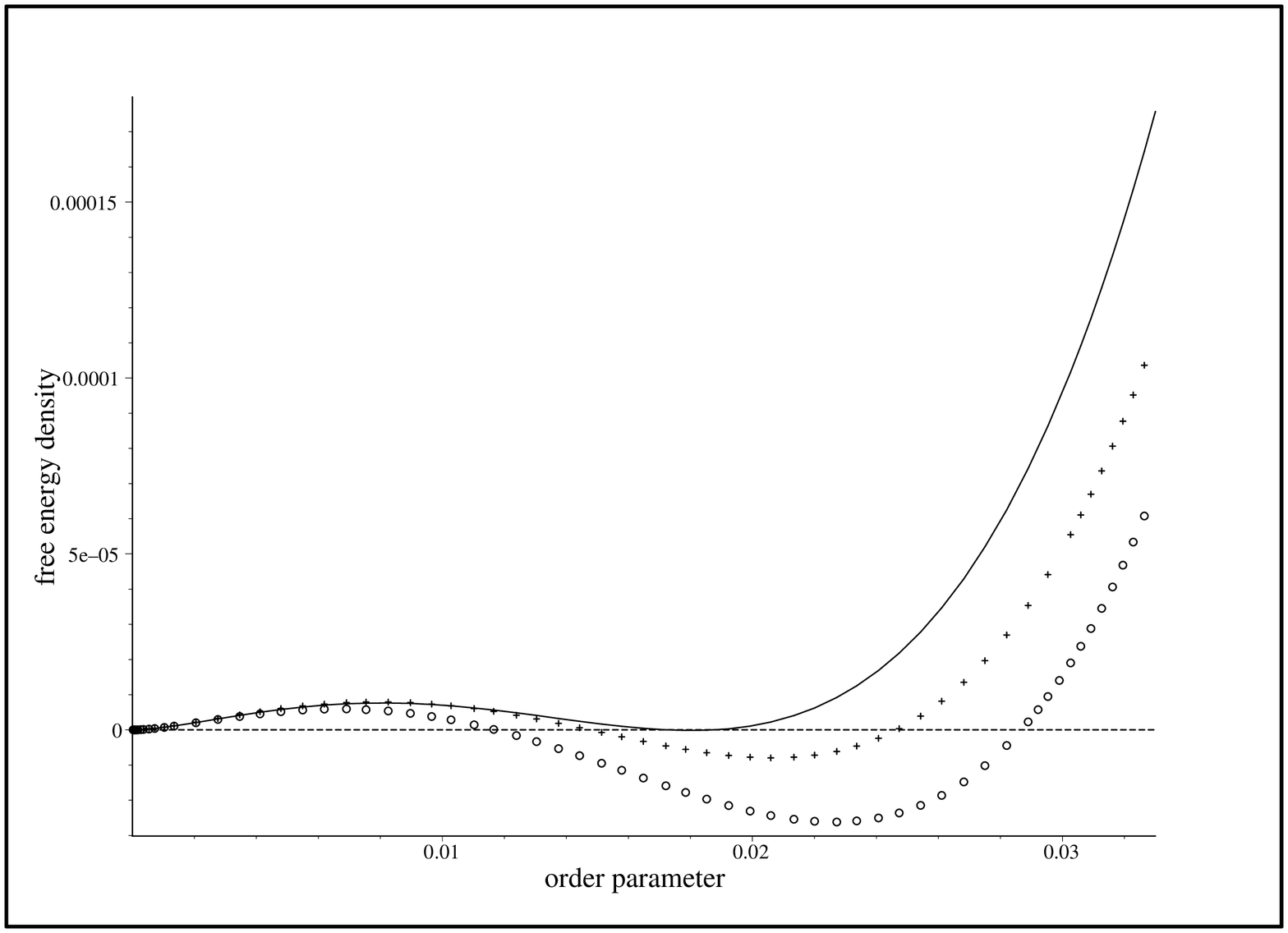,angle=-90, width=12cm}\\
\end{center}
\caption{The function $f(\varphi)$ for Al films of thickness $L_0 =
0.4\mu$m: the solid line corresponds to ET, the line of crosses ($+$)
represents LT, and the line of circles ($\circ$) stands for SLT. All
curves are calculated for $T_{\mbox{\scriptsize eq}}$ corresponding to
ET (see the text).} \label{ST3f1.fig}
\end{figure}

\begin{figure}
\begin{center}
\epsfig{file=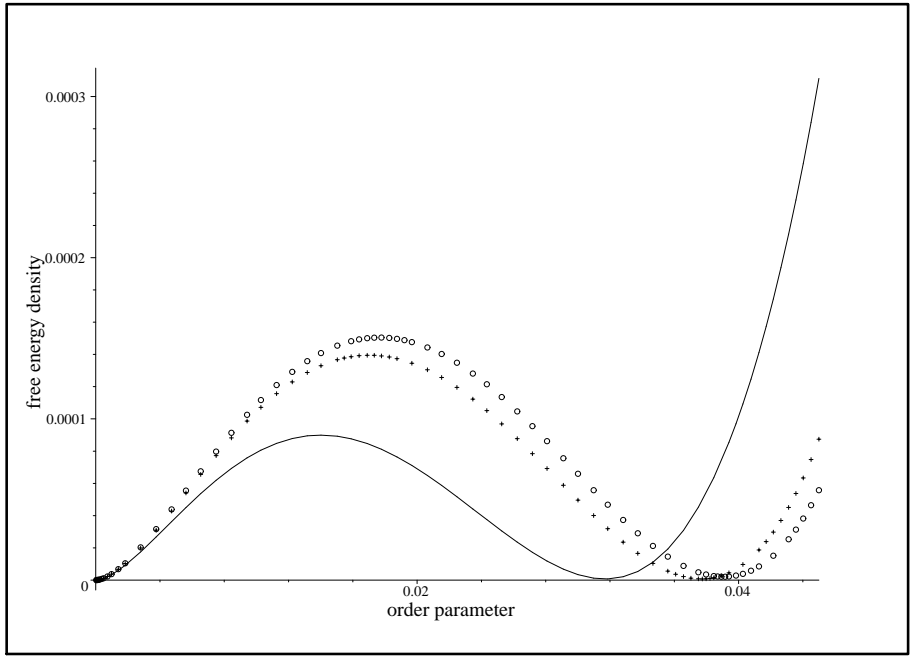,angle=-90, width=12cm}\\
\end{center}
\caption{The function $f(\varphi)$ for Al films of thickness $L_0 =
0.1\mu$m: the solid line (--) corresponds to ET, the line of crosses
($+$) represents LT, and the line of circles ($\circ $) stands for SLT.
All curves are calculated for $T_{\mbox{\scriptsize eq}}$ corresponding
to the respective variant of the theory (see the text).}
\label{ST3f2.fig}
\end{figure}

The shift $t_{\mbox{\scriptsize eq}}$ of the equilibrium transition
temperature due to magnetic fluctuations is very small in both bulk Al
and thin Al films so the difference $(T_{\mbox{\scriptsize eq}}
-T_{co})$ can be neglected in all calculations of thermodynamic
quantities near $T_{\mbox{\scriptsize eq}}$. The equilibrium jump
$\varphi_{\mbox{\scriptsize eq}}$, or, equivalently,
$|\psi|_{\mbox{\scriptsize eq}}$, is one order of magnitude higher in
the film with $L_0=0.1 \mu$m than in bulk Al but the latent heat $Q$ is
$10^2$ times bigger for films. These values are almost one order of
magnitude higher for  $L_0 \sim 0.05\;\mu$m than for $L_0 = 0.1\;\mu$m.
The numerical data in Table 1 are obtained by
 SLT for the bulk Al samples and by  ET for the Al film of
thickness $L_0 = 0.1\;\mu$m; for the abbreviations SLT, LT and ET, see
Sec.~2.4. The difference in the numerical results obtained from ET, LT
and SLT will be discussed in the remainder of the paper.

The investigation of bulk superconductors yields the same results
irrespective of whether we analyze the free energy $f_3(\psi)$ by ET,
LT or SLT. The situation in quasi-2D superconductors is however
different; the three different variants of treatment of the free energy
give different results, in particular, for relatively small thicknesses
($L_0 \ll \xi_0$). This feature of free energy, Eq.~(\ref{eq60}), is
illustrated in Fig.~1, where three curves for three different variants
of $f(\psi)$ are shown for Al films of thickness $L_0 = 0.4\;\mu$m. The
solid line corresponds to ET, the line of crosses represents the LT
result, and the line of circles stands for SLT. All three curves are
calculated for $t_{\mbox{\scriptsize eq}} \equiv
t_0(T_{\mbox{\scriptsize eq}}) = - 0.00057$, which is the ET
equilibrium phase transition temperature $T_{\mbox{\scriptsize eq}} =
0.9994 T_{c0}$. Note, that $\varphi_{\mbox{\scriptsize eq}}$ is the
nonzero global minimum of $f(\varphi)$ and the function $f(\varphi)$
depicted in Fig.~1 has only one minimum for $\varphi >0$ because all
curves are calculated in the thermodynamic regime corresponding to the
stable Meissner phase.

The main conclusion that can be made from Fig.~1 is that the two
variants of the Landau expansion give approximately the same
quantitative results and therefore, the factor $(1+t_0)$
in~(\ref{eq60}) can always be substituted by unity, though the present
investigation is intended to quite small physical effects. This
conclusion is consistent with the argument~\cite{Shopova1:2003} that
allows to use the same approximation $(1+t_0) \approx 1$  for the
calculation of $Q$ and $\Delta C$ in both variants of the theory: ET
and LT. Besides, the Fig.~1 shows that both variants of LT give
slightly higher equilibrium phase transition temperatures
$T_{\mbox{\scriptsize eq}}$ and substantially higher equilibrium jumps
$\varphi_{\mbox{\scriptsize eq}}$ than ET. Thus, using LT for film
thicknesses $a_0 \ll L_0 < 0.1\mu$m one may obtain up to 10 times
higher value of $\varphi_{\mbox{\scriptsize eq}}$ and up to $10^2$
times bigger latent heat $Q$ than the respective values in Table 1. The
problem is whether these higher values predicted by LT are reliable.

Fig.~2 shows the free energy drawn in the three variants (ET, LT and
SLT) of $f(\varphi)$ for Al films of thickness $L_0=0.1\mu$m. In
Fig.~2, the curves $f(\varphi)$ are drawn at their respective
equilibrium phase transition points. The variation in
$t_{\mbox{\scriptsize eq}}$ ($-0.\:00148$ for the solid line,
$-0.\:00115$ for "+" -line and $-0.\:00046$ for $\circ$-line) are of
order of the typical values of $t_{\mbox{\scriptsize eq}}$ itself so
the differences in $t_{\mbox{\scriptsize eq}}$ due to the way of
calculation cannot be neglected. Although for both variants of the
Landau expansion (LT and SLT), the quantity $\varphi_{\mbox{\scriptsize
eq}}$ is again practically the same, the difference in
$t_{\mbox{\scriptsize eq}}$ is more pronounced for smaller thickness of
Al film and moreover, both variants of Landau expansion are not so good
approximation to the result of the exact calculation (ET) as for
$L_0=0.4 \mu$m. The conclusions, we have already drawn from the results
shown in Fig.~1, are completely confirmed by the form of the curves
from Fig.~2 and, moreover, we see that the deviation of the results of
LT from those of ET becomes bigger with the decrease of film thickness
$L_0$.

\begin{figure}
\begin{center}
\epsfig{file=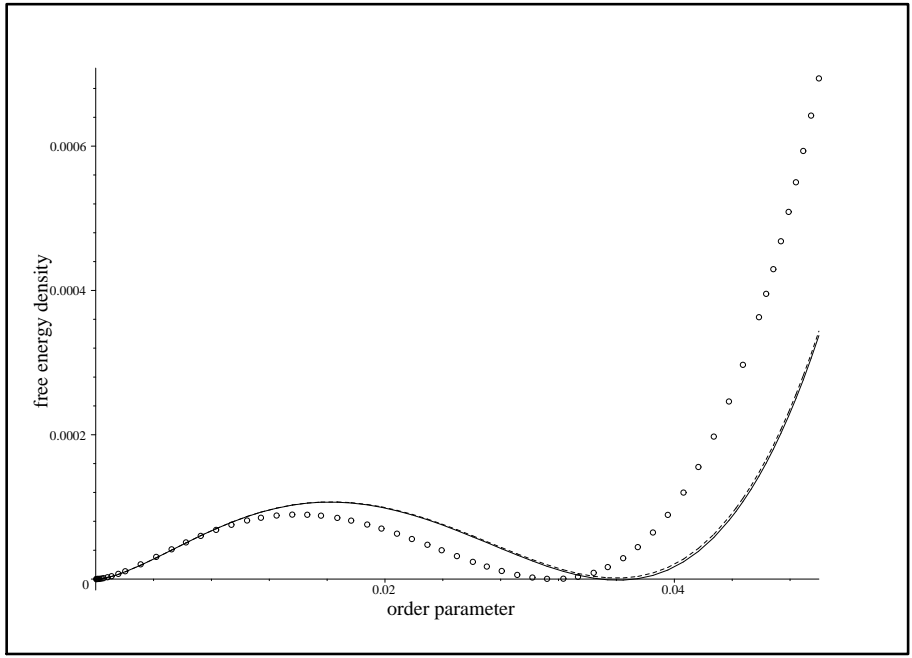,angle=-90, width=12cm}\\
\end{center}
\caption{The function $f(\varphi)$ calculated from the ET for Al films
of thickness $L_0 = 0.1\mu$m and different cutoffs $\Lambda$: the solid
line (--) corresponds to $\Lambda = \pi/\lambda_0$, the dashed line
represents the case $\Lambda = 1/\lambda_0$, and the line of circles
($\circ $) stands for $\Lambda = \pi/\xi_0$. All curves are calculated
at the respective $T_{\mbox{\scriptsize eq}}$ (see the text).}
\label{ST3f3.fig}
\end{figure}

On the basis of the above observations we may conclude that in all
cases when ET and the respective Landau expansions give different
results, the Landau expansion yields a better established first order
transition, with a higher jump $\varphi_{\mbox{\scriptsize eq}}$, and
hence, bigger values of $Q$ and $\Delta C$. In order to establish where
LT is a good approximation we have made systematic numerical
calculations for Al films of different thicknesses $L_0 =0.05 \div
3\mu$m. When $L_0$ is lowered beginning with $3 \mu$m the quantitative
differences between the two variants of theory, with and without Landau
expansion, respectively, become substantial about $L_0 \approx
0.4\mu$m. Bearing in mind the condition for the validity of the Landau
expansion (see, Sec.~2.5) and the requirement for the equilibrium jump
of the order parameter, $\mu \varphi_{\mbox{\scriptsize eq}}^2 <1$ we
may suppose that the predictions done with the help of the Landau
expansions do not satisfy this inequality for  Al films with
thicknesses $L_0 \le 0.1\mu$m. For this relatively small $L_0$-size, ET
is absolutely reliable. The numerical data show that films with $L_0
>0.1 \mu$m are described well quantitatively by the Landau expansion.
The differences between the curves in Figs.~1 and 2 can be neglected in
numerical calculations intended to give theoretical predictions for
experiments.

This general result is supported by the following simple argument. In
the Landau expansion of free energy (\ref{eq60}) for quasi-2D films the
parameter $a$ acquires a cutoff ($\Lambda-$) independent contribution
of the form
\begin{equation}
\label{eq63} \Delta a = \frac{k_BT\rho_0}{4\pi L_0}\:.
\end{equation}
When we compare $\Delta a$ with the bare parameter value
$|a|=(\alpha_0T_{c0}|t_0|)$ for Al, it is easily obtained that $|t_0|$
will not exceed $10^{-4}$ for thicknesses $L_0$ which are of order
$1\:\mu$m or larger. Therefore, for $\L_0 \sim 0.1\:\mu$m the Landau
expansion gives results which are quantitatively different from those
obtained by ET.

We have studied the dependence of the free energy density $f(\varphi)$
of Al films with $L_0 = 0.1\mu$m on the cutoff $\Lambda$. Fig.~3 shows
the free energy density $f(\varphi)$ for three values of the cutoff:
$\Lambda = \pi/\lambda_0$, $\Lambda = 1/\lambda_0$, and $\Lambda =
\pi/\xi_0$. As the cutoff increases from ($\pi/\xi_0)$ to
$\pi/\lambda_0 \sim 10^2(\pi/\xi_0)$, the equilibrium jump
$\varphi_{\mbox{\scriptsize eq}}$ increases, too. We have already
mentioned in Sec.~3.3 that the increase of the cutoff $\Lambda$ for
type I superconductors up to the value $(\pi/\lambda_0)$ is
inconsistent in the present theory. The numerical result for Al films
shown in Fig.~3 is, therefore, a demonstration of the validity of our
arguments about the choice of the cutoff $\Lambda$ presented in
Sec.~3.3. If we take the cutoff $\Lambda \gg (\pi/\xi_0)$ we shall go
beyond the scope of validity of our theory.

There is a similarity between the breakdown of the present theory for
cutoffs $\Lambda \gg (\pi/\xi_0)$ and the breakdown of the condition
($\rho/\Lambda^2) \ll 1$ for the validity of LT at small thicknesses
$L_0$. In both cases, when there is an inconsistency of the theory, we
obtain enhanced values of the characteristic jumps of thermodynamic
quantities at the equilibrium point of the first-order phase
transition.

\section{CONCLUSION}
\label{sec.5.}

We did a detailed analysis of the HLM effect in bulk (3D-)
superconductors and quasi-2D superconducting films within the
self-consistent approximation introduced in Refs.~\cite{Halperin:1974,
 Chen:1978, Coleman:1973}. We have studied for a first time the
validity of this approximation and calculated thermodynamic quantities
of direct experimental interest like the equilibrium jumps of the order
parameter, entropy and specific heat at the point of the
fluctuation-induced first-order phase transition to superconducting
state in a zero external magnetic field. Our investigation is supported
by numerical calculations for bulk Al and Al films.

We have presented for a first time a comprehensive analysis of the
effective free energy of the superconductor in a zero external magnetic
field and on the basis of this analysis we compared the results from
the investigation of the effective free energy without a partial Landau
expansion with those from the Landau expansion of the effective free
energy of quasi-2D films. For quasi-2D films of type I superconductors,
the Landau expansion leads to reliable results, provided the film
thickness is above some value depending on the characteristic lengths
$\xi_0$ and $\lambda_0$, i.e., on the material parameters. For Al films
the results from the Landau expansion become unreliable below film
thicknesses $L_0 \sim 0.1\mu$m. For quasi-2D films of type II
superconductors the Landau expansion of the effective free energy can
be used for any thickness above the level of destruction of
superconductivity $(L_0 \sim 10^{-3}\mu$m).

Our investigation provides a reliable theoretical basis for a future
experimental search of the HLM effect in thin films of type I
superconductors, where the effect is much stronger than in bulk
materials. In accord with preceding works~\cite{Folk:2001,
Shopova:2002, Shopova1:2003, Shopova2:2003, Shopova3:2003} we have
justified earlier results which indicate that the HLM effect will be
better pronounced in films of materials with relatively high values of
the critical magnetic field $H_{c0}$ and relatively small thicknesses
$L_0$. We cannot be certain which superconducting material provides the
best experimental conditions for transport or caloric measurements of
the jumps of the thermodynamic quantities at the point of the
fluctuation-induced first order transition, but from the data for Al, W
and In available from recent studies~\cite{Shopova1:2003,
Shopova2:2003} and the present paper, the most suitable substance seems
to be Al. But investigations of other superconductors may put forward
materials which are even better candidates for the experimental test of
the HLM effect.

Looking for the most convenient material for the experimental search of
the HLM effect we should have in mind a number of other experimental
requirements which are not related to the results from the present
theoretical investigation. Here we shall briefly discuss the problem of
the external magnetic field and the possible change
 of the
superconductivity from type I to type II with the decrease of the film
thickness~\cite{Huebener:2001}. This change for film thicknesses $L_0$,
which are convenient for the experimental study of the HLM effect, is
not a great problem because we have shown in our investigation that the
effect could be observed also in films of type II superconductors,
provided the external magnetic field $H_0$ is very low so the effect of
the vortex phase and the magnetic energy jump ($H_0^2/8\pi)$ at the
phase transition point is negligible. The magnetic energy jump
($H_0^2/8\pi$) may obscure the HLM effect on the latent heat also in
type I superconductors and, therefore the experimental problem for the
elimination of the residual laboratory external magnetic field $H_0$ is
common for both type I and type II superconducting films. If we take as
a basis the latent heat of order~1~[erg/cm$^3$] in Al films with $L_0
\sim 0.1\mu$m, as reported in Sec.~4, the magnetic field which ensures
the ratio $(H_0^2/8\pi Q) \ll 1$ will be obviously about 1~Oe. In
thinner films of convenient materials this experimental condition may
become $H_0 \sim 10$ Oe but no more.

{\bf Acknowledgements.} The authors thank Prof. Dimo I. Uzunov for
useful discussions. Grants from Scenet (Parma) and JINR (Dubna) are
acknowledged.

\end{document}